\PassOptionsToPackage{dvipsnames}{xcolor}
\documentclass[10pt,sigconf,nonacm,screen]{acmart}

\usepackage[all]{nowidow}
\usepackage{xcolor}
\usepackage{subcaption}
\usepackage{hyperref}
\usepackage{acronym}

\usepackage[capitalise,noabbrev]{cleveref}
\crefformat{section}{\S#2#1#3} 
\crefformat{subsection}{\S#2#1#3}
\crefformat{subsubsection}{\S#2#1#3}
\crefrangeformat{section}{\S#3#1#4 to~\S#5#2#6}
\crefrangeformat{subsection}{\S#3#1#4 to~\S#5#2#6}
\crefrangeformat{subsubsection}{\S#3#1#4 to~\S#5#2#6}

\begin{document}


\author{Tim C. Rese}
\affiliation{%
    \institution{TU Berlin \& ECDF}
    \city{Berlin}
    \country{Germany}}
\email{tr@3s.tu-berlin.de}
\orcid{0009-0008-0185-8339}

\author{Nils Japke}
\affiliation{%
    \institution{TU Berlin \& ECDF}
    \city{Berlin}
    \country{Germany}}
\email{nj@3s.tu-berlin.de}
\orcid{0000-0002-2412-4513}

\author{Sebastian Koch}
\affiliation{%
    \institution{TU Berlin \& ECDF}
    \city{Berlin}
    \country{Germany}}
\email{sko@3s.tu-berlin.de}
\orcid{0009-0003-2296-5305}

\author{Tobias Pfandzelter}
\affiliation{%
    \institution{TU Berlin \& ECDF}
    \city{Berlin}
    \country{Germany}}
\email{tp@3s.tu-berlin.de}
\orcid{0000-0002-7868-8613}

\author{David Bermbach}
\affiliation{%
    \institution{TU Berlin \& ECDF}
    \city{Berlin}
    \country{Germany}}
\email{db@3s.tu-berlin.de}
\orcid{0000-0002-7524-3256}

\title{Increasing Efficiency and Result Reliability of Continuous Benchmarking for FaaS Applications}



\begin{abstract}
    In a continuous deployment setting, Function-as-a-Service (FaaS) applications frequently receive updated releases, each of which can cause a performance regression.
    While continuous benchmarking, i.e., comparing benchmark results of the updated and the previous version, can detect such regressions, performance variability of FaaS platforms necessitates thousands of function calls, thus, making continuous benchmarking time-intensive and expensive.
		
    In this paper, we propose DuetFaaS, an approach which adapts duet benchmarking to FaaS applications.
    With DuetFaaS, we deploy two versions of FaaS function in a single cloud function instance and execute them in parallel to reduce the impact of platform variability.
    We evaluate our approach against state-of-the-art approaches, running on AWS Lambda.
    Overall, DuetFaaS requires fewer invocations to accurately detect performance regressions than other state-of-the-art approaches. In $98.41\%$ of evaluated cases, our approach provides equal or smaller confidence interval size. DuetFaaS achieves an interval size reduction in $59.06\%$ of all evaluated sample sizes when compared to the competitive approaches.
\end{abstract}

\maketitle

\section{Introduction}
\label{sec:introduction}

Function-as-a-Service (FaaS) is an established cloud computing paradigm that shifts operational responsibilities to the cloud provider and lets developers focus on implementing business logic~\cite{bermbach2021future,li2022serverless,hendrickson2016serverless}.
In addition to elastic scalability, FaaS customers also benefit from pay-per-use billing where only the time a function actually runs is billed.
This makes continuous observation of FaaS function performance a necessity, as any performance regression in a function will not only be visible to end users but will also be directly reflected in the cloud service bill.
In a continuous integration and deployment (CI/CD) setting, where new function releases are deployed frequently, this is even more important as new code is likely to cause performance regressions~\cite{grambow2022microbenchmark}.
To detect such performance regressions early before they can affect the live application, both van Hoorn et al.~\cite{van2012kieker} and Grambow et al.~\cite{grambow2019continuous} have proposed to benchmark new releases as part of the CI/CD pipeline -- continuous benchmarking.

The naive approach to continuous FaaS benchmarking, i.e., deploying two versions of a function and iteratively invoking both, requires hundreds to thousands of function invocations in order to yield meaningful results~\cite{book_bermbach2017_cloud_service_benchmarking,leitner2016patterns,bermbach2017quality,schirmer2023nightshift}.
This is simply a result of the high level of abstraction of FaaS platforms, which build on complex hardware and software stacks, and are thus subject to performance variability~\cite{scheuner_crossfit,scheuner_phdthesis}. 

Previous research has found that variability differs across FaaS platforms, with performance differences up to $15\%$ being detected~\cite{schirmer2023nightshift}. It also has proposed methodologies and strategies to reduce and mitigate the impact of such influences~\cite{grambow2023efficiently,grambow2019continuous,grambow2021befaas}.
Thus far, all strategies use either parallel, independent function invocations, which can be affected by performance variability between the machines the different function instances are deployed on, or sequential function invocations within the same function instances, which is subject to temporal performance variability.
We instead propose adapting the concept of duet benchmarking~\cite{bulej2020duet} from microbenchmarks to FaaS applications: both function versions are deployed in a \textit{shared} cloud function instance and are executed in \textit{parallel}, with each function being provided the same amount of resources.
This minimizes influence of temporal performance variation and heterogeneous underlying hardware.
In this regard, we make the following contributions:

\begin{itemize}
    \item We propose \emph{DuetFaaS}, a continuous benchmarking approach for integration into CI/CD pipelines that adapts duet benchmarking to FaaS environments (\cref{sec:approach}).
    \item We implement an open-source proof-of-concept prototype of DuetFaaS, using AWS Lambda as an example platform (\cref{sec:experiment:prototype}).
    \item We evaluate the DuetFaaS approach in experiments, comparing it to the traditional independent function invocation and state-of-the-art sequential invocation within a shared function instance (\cref{sec:experiment}).
\end{itemize}

Through our experiments, we show that DuetFaaS provides equal or smaller confidence interval sizes in $98.41\%$ of experiments and requires as few as 100 function invocations to receive results with a confidence interval with a width of less than 2 percent of the median.
Compared to the state of the art, this means that DuetFaaS provides almost always more stable results with less repetitions, i.e., is faster and less expensive.
\section{Current State of Function-as-a-Service (FaaS) Release Benchmarking}
\label{sec:background}
In this section, we highlight the key concepts required to understand the current field of FaaS benchmarking, and provide related work where alternative approaches are discussed.

\subsection{Function-as-a-Service (FaaS)}

FaaS is a popular event-driven model in which developers split their application into functions and host them in a cloud environment~\cite{paper_bermbach2021_cloud_engineering}.
The functions are then able to run based on various triggers, such as events or HTTP calls.
From a developer perspective, FaaS is attractive as it offers a pay-per-use model (rather than the pay-as-provisioned model of traditional cloud infrastructure services) with scale-to-zero and because the entire function lifecycle including failure management and scaling is handled by the platform provider~\cite{bermbach2020using}.

\subsection{Continuous Benchmarking}

\begin{figure}
    \centering
    \includegraphics[width=0.4\textwidth]{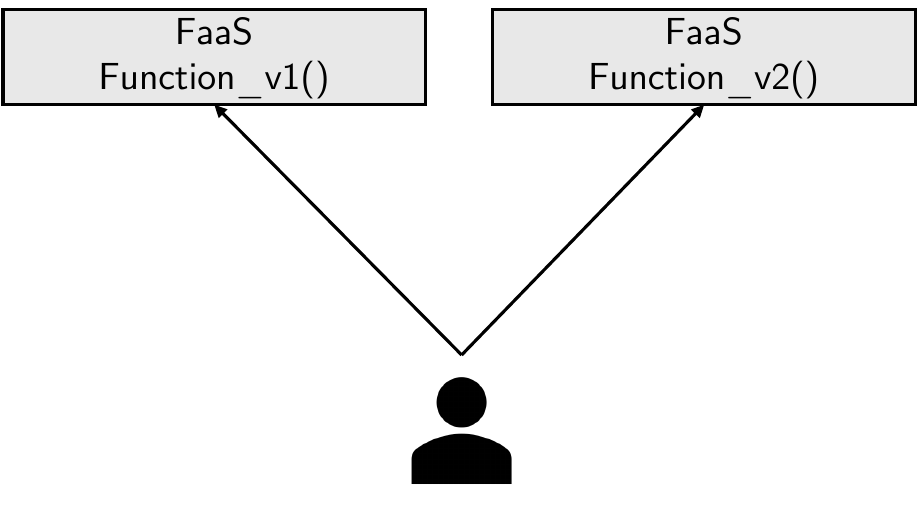}
    \caption{Traditional FaaS Benchmarking, where function versions are deployed independently of one another. Load generation calls are then made separately to both.}
    \label{img:trad}
\end{figure}

Continuous benchmarking aims to include a benchmarking step within a deployment pipeline the same way tests are commonly included. This means that every time an application is updated, a benchmark is run to detect a possible performance change. While performance is a non-functional requirement, it has a direct impact on cost. Benchmarks aim to make this performance change visible, which can help decision making when evaluating an update.
Several works have called for performance change detection directly after a change is published~\cite{japke2023early,javed2020perfci,leitner2016patterns, grambow2022microbenchmark}.
Other recent work has also highlighted the importance of benchmarking during the deployment process, with approaches showing potential to detect significant performance regressions before final deployment of a release~\cite{japke2023early,grambow2023efficiently,grambow2019continuous}.

\subsection{Continuous FaaS Application Benchmarking}
\begin{figure}
    \centering
    \includegraphics[width=0.3\textwidth]{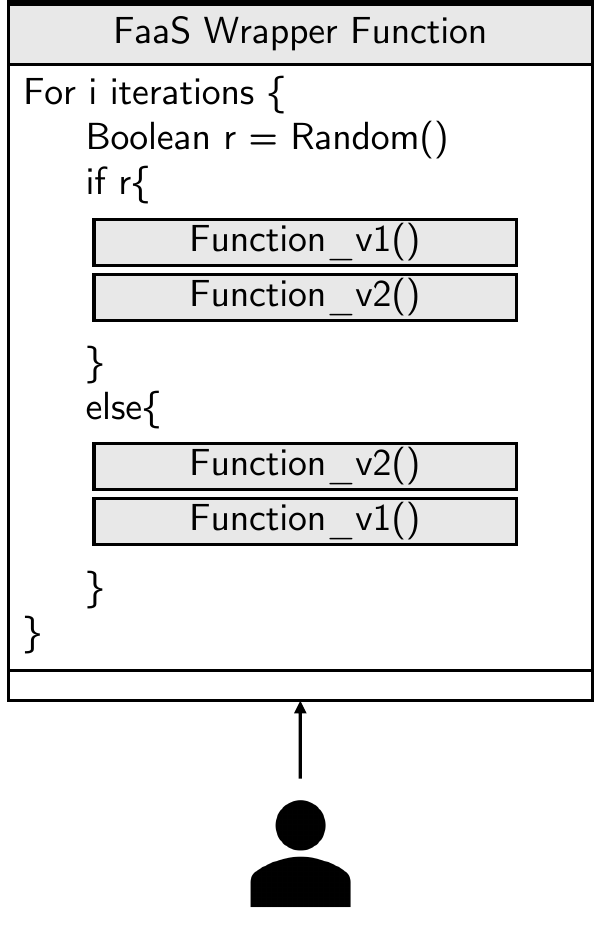}
    \caption{RMIT Functionality within faasterBench, where a wrapper artifact containing both versions determines a random factor. This random factor decides in which order the functions are run, and the experiment is repeated for several trials.}
    \label{img:rmit}
\end{figure}
Continuous benchmarking is essential for FaaS applications, as these are developed in the same way other modern applications are. The traditional strategy to benchmark such applications is to deploy both version artifacts separately of one another. Each version, as shown in \cref{img:trad}, is run on separate instances and no form of resource sharing is required. An artificial load is generated by a benchmarking client, and then sent to both individually. This approach inherently includes several aspects that can be improved. As various uncontrollable factors exist in the cloud, these may impact your artifacts differently. Additionally, the underlying executing instance may vary, which can also influence overall result quality~\cite{book_bermbach2017_cloud_service_benchmarking,paper_bermbach2017_expect_the_unexpected}.

Grambow et al. improved upon the traditional independent approach by implementing the randomized multiple interleaved trials (RMIT) methodology. In RMIT, both artifacts are hosted on the same instance. A random factor decides the execution order of the function versions, which are then run. This is repeated for multiple trials, with the execution order being determined randomly before every iteration~\cite{abedi2017conducting,grambow2023efficiently}. This provides several benefits, with the key benefit being that the executing instance is now the same. Additionally, it ensures that execution order has no impact on the results, as the order is randomly determined every iteration.

\cref{img:rmit} highlights how Grambow et al. implemented the RMIT methodology in previous research. The authors showed that RMIT can improve result accuracy and requires fewer results than the traditional approach, where both versions are deployed separately and called independently of one another~\cite{grambow2023efficiently}.

Duet benchmarking has however not been implemented for FaaS application releases to the best of our knowledge. Duet benchmarking aims to guarantee resource fairness for both function versions by executing them in parallel on the same instance.
The artifacts are usually containerized, provided the same amount of resources (CPU, memory, etc.), and then executed in parallel, which also causes occurring uncontrollable factors to influence both versions equally~\cite{bulej2020duet}.
This process enables a fair benchmarking procedure when wanting to efficiently detect performance variation in software application releases.
Duet benchmarking was originally introduced to improve measurement accuracy on cloud instances, but has been extended to other fields~\cite{bulej2020duet}.
More recently it was shown that it is well suited for microbenchmarks, and early findings also showed accuracy improvements with benchmarking suites~\cite{bulej2020duet,bulej2019initial}.

However, published work includes its introduction and evaluation in general cloud benchmarking~\cite{bulej2020duet}, while other papers consider it as a valid remediation step to possibly improve their results~\cite{schirmer2023nightshift}.

\subsection{Result Analysis}

Performance Change Detection for application releases includes the benchmarking and evaluation of release versions to detect noticeable performance variations. This process is done to ensure that both the user and developer do not experience unexpected side effects.
The user could expect an answer from the application within a certain time period, while the developer has to pay for the additional duration both in local and FaaS environments.

Bootstrapping Percentile Intervals involves the re-sampling of collected values with replacement to calculate confidence intervals for a collection of measurements.
A percentile confidence interval is a range that contains a specified percentage of the data, defined by a minimum and maximum value.
The standard procedure to create a \textit{x}\% percentile interval is to sort the results, and remove ($\textit{x}/2$)\% of values from both the top and bottom end. 
The smallest and largest value are then the interval bounds.

We rely on this methodology in our study as it is commonly used and makes no assumptions about the actual distribution of measurements~\cite{puth2015variety,kalibera2020quantifying}.

\section{Duet Benchmarking FaaS Application Releases}
\label{sec:approach}

The duet benchmarking methodology requires certain conditions to be met so it can be used. Additionally, as we want to implement the methodology for FaaS applications in modern deployment pipelines, other guarantees must be given. In this section, we discuss what requirements exist and how our design meets these.

\subsection{Parallel Isolated Execution}
FaaS platforms do not natively support the duet benchmarking methodology, as the degree of control required to run such experiments is higher than what such a service usually provides. 
Duet benchmarking requires functions to be run in parallel on the same machine, share the available memory, but be isolated to their own CPU core. 

The amount of factors that we need to control exceeds that of a native deployment, hence is why we made the choice to deploy a web server of sorts as a FaaS function.
This allows us to pass configuration options to the server, which ensure both the parallel execution of both artifacts, and allowing us to isolate each artifact to a separate core.
We can use the primary process running on our server to coordinate worker processes, which can then be told to run each artifact in parallel.
We can also use the same primary process directly as the scheduler. Once both worker processes are ready, we can coordinate the parallel execution of function artifacts by synchronously starting the artifacts on the workers. 

\subsection{Realistic Environment}
We can only compare duet benchmarking in FaaS with other state-of-the-art approaches if we also run it in a realistic cloud environment. 
A local deployment lacks the uncontrollable factors that the cloud is subject to, as well as being difficult to include in a deployment pipeline with continuous benchmarking.
Another option would be to host the function on a cloud instance, but not as a FaaS function. This would be closer to what we are trying to achieve, but performance would likely still differ from what one could expect in an actual FaaS environment. 

Our design solves this by being deployable as a FaaS function to a public cloud provider.
This includes the benefit of being usable in a production FaaS environment, which also makes our experiment results more comparable with other approaches. We believe this to the most "FaaS-like" environment possible when wanting to run duet benchmark experiments. 

\subsection{Pipeline Integration}
As we want duet benchmarking to be used as a continuous benchmarking step in deployment pipelines, we need to set up our approach to be easily includable in such tools. 

As our function is deployable to a public cloud platform, we can also use an action within a deployment pipeline to run fully automated experiments. 
We can set up the action to monitor the version control system for relevant changes, which then trigger an automatic deployment, benchmark execution, and destruction of duet benchmark experiments. 
This is also another reason why a local deployment would not suffice, as it would prove difficult to make every possible local environment compatible with a deployment pipeline. 

Grambow et al. deployed multiple function artifacts in parallel to more quickly complete benchmark experiments, as well as reduce the impact that a 'bad' instance can have on the overall results. As we want experiments to quickly finish, we also make our approach deploy multiple FaaS instances at once, each hosting our duet benchmarking approach. This provides the same benefits as in other research, both reducing the impact of 'bad' instances and reducing the required time for benchmark completion~\cite{grambow2023efficiently}.

\begin{figure*}
    \centering
    \includegraphics[width=\textwidth]{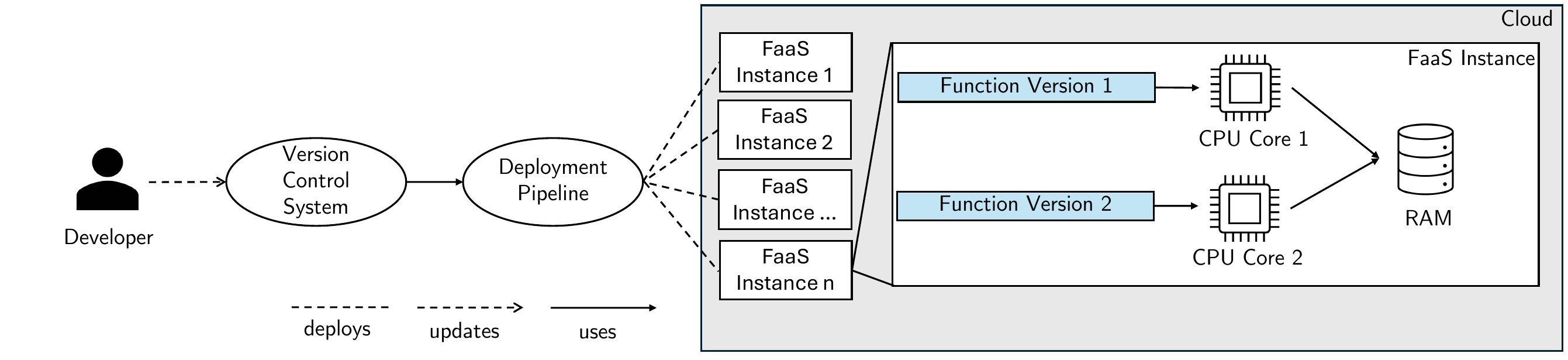}
    \caption{Whenever a change in the repository is made, a deployment pipeline uploads the updated and previous version to multiple FaaS instances. Here, both versions are run on separate cores and in parallel to enable duet benchmarking.}
    \label{img:systemdesign}
\end{figure*}

\cref{img:systemdesign} shows the simplified design architecture of DuetFaaS (the name of our approach). 
\section{Evaluation}
\label{sec:experiment}
We evaluate the DuetFaaS approach by implementing a proof-of-concept prototype for AWS Lambda (\cref{sec:experiment:prototype}). We design an experiment to evaluate our prototype (\cref{sec:experiment:design}), along with comparing it to independent invocations and state-of-the-art randomized sequential invocations (\cref{sec:experiment:ex1}). We also evaluate the impact of sample sizes, i.e, we regard how the number of included experiment results has an influence on the confidence interval size (\cref{sec:experiment:ex2}).
\subsection{Proof-of-Concept Prototype}
\label{sec:experiment:prototype}
We build our prototype in AWS and rely on a repository built specifically for AWS\footnote{\url{https://github.com/awslabs/aws-lambda-web-adapter}}.
This allows us to deploy a web server to Lambda, which can host our function artifacts.
We can configure this FaaS function to run an Express server in cluster mode.
This creates a master process, which creates two worker processes with each of the workers being run as a separate execution thread. Once each worker is ready, we use the master process to synchronously call the function executions and enable duet benchmarking within an actual FaaS environment.

Our prototype is already CI/CD compatible and can be triggered from GitHub Actions, we however make slight adaptations to the normal functionality for our experimental evaluation. Our prototype is deployed independently of a deployment pipeline, so we can evaluate the approach with repeated experiments.

Our benchmarking tool is open-source and available on GitHub, along with a detailed explanation how to run one's own experiments\footnote{\url{https://github.com/timchristianrese/DuetFaaS-code}}.

\subsection{Experiment Design}
\label{sec:experiment:design}
\begin{figure*}
    \begin{subfigure}{0.5\textwidth}
        \centering
        \includegraphics[width=0.99\textwidth]{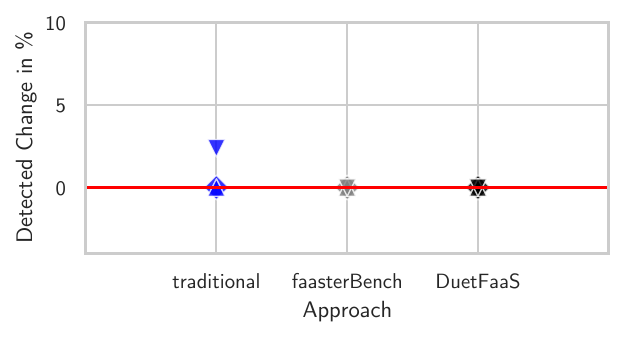}
        \caption{\texttt{CPU} A/A Experiment}
    \end{subfigure}
    \begin{subfigure}{0.5\textwidth}
        \centering
        \includegraphics[width=0.99\textwidth]{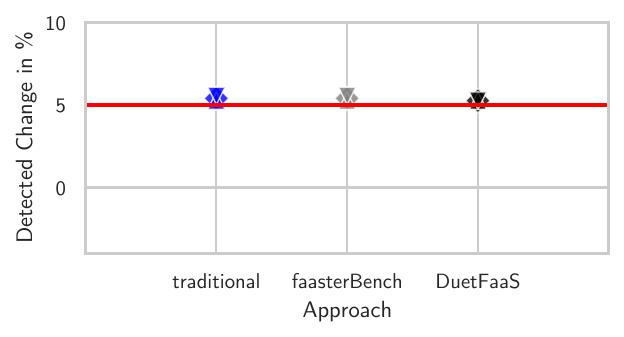}
        \caption{\texttt{CPU} A/B Experiment}
    \end{subfigure}
    \begin{subfigure}{0.5\textwidth}
        \centering
        \includegraphics[width=0.99\textwidth]{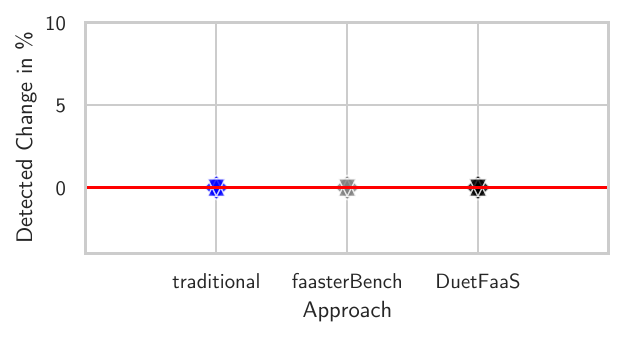}
        \caption{\texttt{MEM} A/A Experiment}
    \end{subfigure}%
    \begin{subfigure}{0.5\textwidth}
        \centering
        \includegraphics[width=0.99\textwidth]{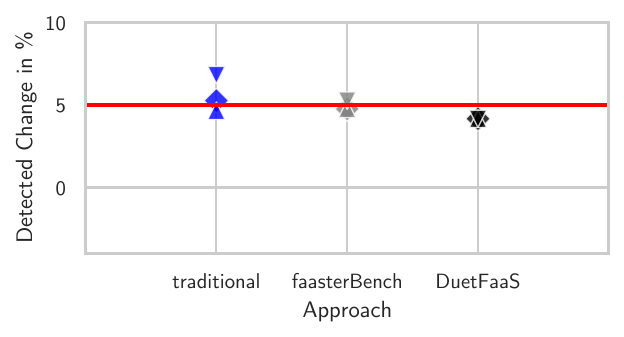}
        \caption{\texttt{MEM} A/B Experiment}
    \end{subfigure}
    \caption{Performance regression detected by each approach in 1,500 benchmark repetitions, along with the $99\%$ confidence interval. Red lines mark the `true' performance regression of the function. Duet benchmarking provides accurate intervals compared to the other approaches. It however slightly misses the true regression in one of our configurations.}
    \label{fig:ci1500sizes}
\end{figure*}
Our experiments with DuetFaaS aim to show to what extent our approach can improve the accuracy of detecting performance regressions in FaaS functions.
To this end, we deploy two versions of a FaaS function, execute performance benchmarks against that function using different benchmarking approaches, and then compute the size of the bootstrapped results' confidence intervals.
Additionally, we are interested in how many benchmark executions are necessary to achieve that confidence interval.
Ideally, we reduce both the confidence interval size (meaning results are more accurate) while also reducing the number of necessary executions (meaning the performance evaluation is cheaper and faster to execute).
Naturally, there is a trade-off between measuring more data for a smaller confidence interval and reducing the number of benchmark execution, which is why we measure confidence interval size with a varying number of benchmark executions included in the result computation, evaluating the compared approaches along this trade-off.
We deploy two functions to AWS Lambda, a CPU-intensive neural network mutation~(\texttt{CPU})~\cite{mlflappy} and a memory-intensive implementation of the \emph{Sieve of Eratosthenes} algorithm used to calculate prime numbers~(\texttt{MEM})~\cite{sieve}.
Both functions are implemented in JavaScript.
Each function is allocated 2048 MB of memory for the traditional and randomized sequential approach.
In our DuetFaaS approach, we allocate 3008 MB of memory to the function, as we need to run two artifacts in parallel. 
This amount of memory allocates at least two cores to our Lambda function, although less would have sufficed~\cite{awsyoutube}.

We perform A/A and A/B benchmarks of each function.
For A/A benchmarks, we `compare' two identical implementations of the function, which ideally yields no detected performance regression.
In our A/B benchmarks, we artificially inject an artificial performance regression in each function by modifying the code accordingly, i.e., performing $5\%$ more mutations in the \texttt{CPU} function and increasing the prime number upper threshold by $5\%$ in the \texttt{MEM} function.

We compare the DuetFaaS approach against an improved traditional independently called benchmark and sequential randomized (RMIT / faasterBench) invocation.
In the improved independent approach, we invoke each function version individually without the other version. 
However, the versions are not deployed individually, and instead are put in a wrapper artifact.
This leads to the executing instance being the same, which would very likely not occur if we deployed both versions completely independent of one another.
Sequential randomized invocation is adapted from Grambow et al.~\cite{grambow2023efficiently}.
Here, both function versions are deployed together and invoked sequentially within the same function executor. 
A wrapper decides the execution order for each invocation, implementing an RMIT strategy. This is then repeated for multiple iterations, with the order being randomly determined each iteration.
We run 1,500 benchmark repetitions for each approach, calculate the median performance change for each benchmark, and calculate a confidence interval with a confidence level of $99\%$ using bootstrapping for the performance changes.
To assert reproducibility, we repeated all experiments three times across multiple times-of-day~\cite{schirmer2023nightshift} and observed very similar results.
Further, we note that we remove cold start executions from our results.
\begin{figure*}
    \begin{subfigure}{0.5\textwidth}
        \centering
        \includegraphics[width=0.99\textwidth]{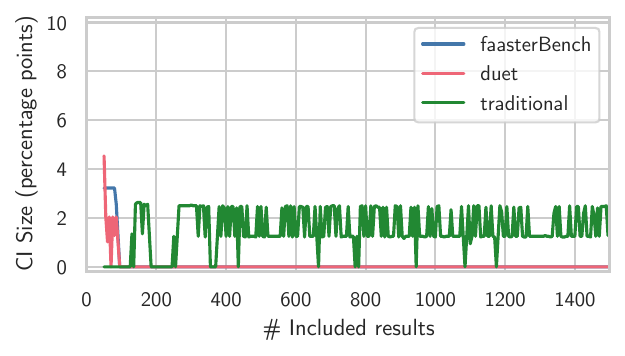}
        \caption{\texttt{CPU} A/A Experiments}
    \end{subfigure}%
    \begin{subfigure}{0.5\textwidth}
        \centering
        \includegraphics[width=0.99\textwidth]{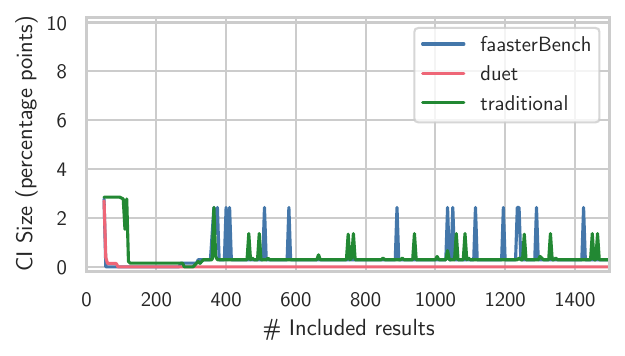}
        \caption{\texttt{CPU} A/B Experiments}
    \end{subfigure}
    \begin{subfigure}{0.5\textwidth}
        \centering
        \includegraphics[width=0.99\textwidth]{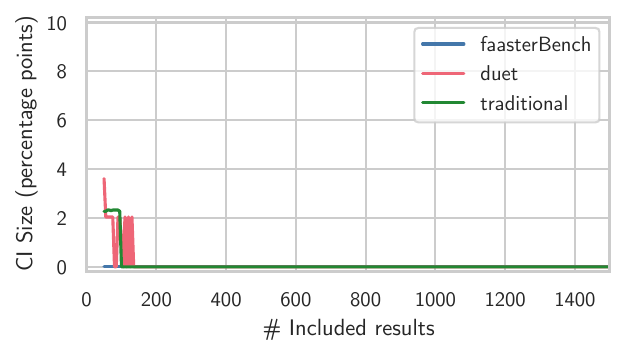}
        \caption{\texttt{MEM} A/A Experiments}
    \end{subfigure}%
    \begin{subfigure}{0.5\textwidth}
        \centering
        \includegraphics[width=0.99\textwidth]{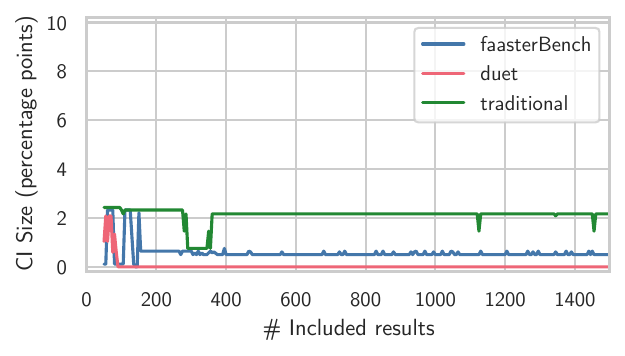}
        \caption{\texttt{MEM} A/B Experiments}
    \end{subfigure}
    \caption{Duet Benchmarking consistenly produces extremely accurate intervals when only including 100 results, which the other approaches do not achieve even with 1,500 repetitions in some configurations.}
    \label{fig:cisizes}
\end{figure*}
\subsection{Performance Regression Detection}
\label{sec:experiment:ex1}

We will focus on one of our experiment runs in this analysis, however the other runs produced similar results. In \cref{fig:ci1500sizes}, we show the performance regression that each approach detects with 1,500 made calls, i.e, from one of our experiment runs.

Our experiments show that duet benchmarking detects the performance change better than the compared approaches regardless of function type and regression injection.
In all benchmarked configurations, duet benchmarking provides smaller confidence intervals, as well as more accurately detecting the actual 'true' regression when running 1,500 experiment repetitions.

Across all four experiments, duet benchmarking's average CI size is 0.01 percentage points when including 1,500 benchmark experiments, which is \ $99.19\%$ / $94.74\%$ smaller than both the traditional independent invocation (1.24) and the randomized sequential one (0.19).
Therefore, we recommend the duet benchmarking methodology for benchmarking FaaS applications when running longer experiments.

We find the \texttt{CPU} function to provide significant improvements over currently used approaches.
Both configurations provided an interval size of 0.0 percentage points when using duet benchmarking, compared to the 0.15 of randomized sequential and 1.40 of traditional independent.
This highlights the staggeringly high level of accuracy that duet benchmarking is able to provide, even in FaaS environments.

Our \texttt{MEM} function provides similar results.
We again see a significant improvement from the other approaches to duet benchmarking, with interval size reductions of approx.\ $96.15\%$ and $99.28\%$ compared to the randomized sequential and traditional independent methodology, respectively.
Our results are also close to the true injected regression, showing that duet benchmarking is viable as a FaaS benchmarking methodology.

\subsection{Impact of Sample Size}
\label{sec:experiment:ex2}
When running duet benchmarking experiments in a deployment pipeline, we optimally want experiments to be short and accurate.
We therefore evaluate the impact of sample size on the overall confidence interval to see if a similarly accurate result could be achieved when running fewer experiments.
We do this by reducing the number of included experiment results, and regarding the effect on confidence interval size.

In \cref{fig:cisizes}, we show how the confidence interval size develops as the sample size increases from 50 to 1,500 samples in steps of 5, i.e., as we perform more experiments for each benchmarking strategy.
We do not evaluate smaller sample sizes, as bootstrapping has been found to underestimate the interval size when including only few samples~\cite{linnet2000nonparametric}.

Our experiments show that a similar level of interval size is already reached with as few as 100 results (depending on the configuration), with all scenarios requiring less than 200 calls made to achieve such a value.
This is noticeably different to the other approaches, which require more calls before a stable accuracy level is achieved.
When only including 100 results, duet benchmarking already provides an average interval size of only 0.26 percentage points, while the traditional independent provides a size of 1.29.
The randomized sequential methodology is better than the traditional at this point, with an interval of over 0.28 percentage points.
We notice large improvements by both the randomized sequential and duet approaches compared to the traditional approach. Of note is duet benchmarking's reduced interval size over both approaches in A/B configurations. 
Worthy of note is that duet benchmarking is commonly able to provide narrower CIs with fewer repetitions than comparing results with more.
Considering the noticeable accuracy stagnation in the traditional and randomized sequential methodology leads us to conclude that these approaches would likely never reach the level of accuracy that duet benchmarking is able to achieve with only 100 results, depending on the configuration.

\subsection{Key Findings}
Duet benchmarking provides narrower confidence intervals than both the traditional independent and randomized sequential approach by a significant margin.
We find that applying duet benchmarking can reduce interval size by up to $99.28\%$ depending on the experiment configuration, along with working well for varying function types and being able to detect the injected performance regression accurately.
Additionally, duet benchmarking requires fewer results to provide an accurate and stable confidence interval.
While other approaches require up to 500 results to provide a stable confidence interval size, duet benchmarking only needs 100.

Overall, duet benchmarking provides equal or better interval sizes than the other approaches in $98.41\%$ of evaluated cases, and provides smaller sizes with fewer results than other approaches with more repetitions.
We propose to include duet benchmarking as a continuous benchmarking step in deployment pipelines.

\section{Discussion}
\label{sec:discussion}
In this section, we discuss the meaning of our findings and how they impact FaaS benchmarking.

Duet benchmarking can be used as a valid alternative to accurately detect a possible change in performance for FaaS applications and we propose to include it as a continuous benchmarking step within a deployment pipeline.

Our experiments show that duet benchmarking provides accurate results with as few as 100 results and does not necessarily require more invocations, as a high degree of result accuracy is already reached with that many repetitions.
This indicates that a short benchmarking step is already enough to provide a usable result for the FaaS application, which can help speed up the deployment process.
While we consider duet benchmarking to already be usable for specific FaaS applications, further improvements and implementations must be made to extend duet benchmarking to other FaaS environments.

\subsubsection*{Measurement of Execution Time}
The way we currently measure the time in duet benchmarking and how we do it for the traditional and RMIT approach slightly differs.
We take the actual required CPU time for duet benchmarking, while our two other approaches record timestamps before and after the function call which we then calculate the difference of.
While we suspect that changing the measurement style to actual CPU time would slightly improve the results of the other approaches, it is unlikely to impact our conclusion. We suspect that duet benchmarking likely still provides more accurate results than the compared approaches.

\subsubsection*{Applicability}
Duet benchmarking only provides insights into the relative performance change of function artifacts and should not be used for cost prediction. This is due to the nature of our approach, as available memory and CPU resources are split across two functions. In an actual production environment, the artifact would be able to access all available resources, which will likely benefit its execution duration. 

This is however not the problem a developer wants to address when evaluating separate application releases. The focus here usually lies on the relative performance change to other versions and not on the actual performance.

Future work aims to create a separate deployment step, where both function artifacts are deployed 'normally'. This would allow for our approach to also cover a cost prediction aspect.

Alternatively, other benchmark suites already exist which cover this feature~\cite{grambow2021befaas, grambow2023efficiently, copik2021sebs}.

\subsubsection*{Ease of Use}
DuetFaaS is easily includable in GitHub's deployment pipeline, but still requires additional overhead such as setting up the web server and likely small adaptations to the benchmark artifacts when comparing it to the traditional approach, where the original version is simply benchmarked.
When evaluating one's application, a trade-off between implementation simplicity and result reliability must be made by the developer.

Given the large accuracy benefit that we find with duet benchmarking, however, we consider this trade-off to be worthy of the effort. 
Additionally, the required overhead can be completed relatively quickly depending on the complexity of the FaaS application, and good task separation within your functions will simplify this process.

\subsubsection*{Limited Providers}
As of now, we only implemented the duet benchmarking methodology within AWS, and have not validated our approach on other platforms. While we believe that other platforms would provide similar tendencies, we have not verified this and are currently working on implementing our approach on other platforms.

We are currently working on implementing our approach within other cloud providers, and aim to make it as generally applicable to make it compatible with almost any cloud environment.
\subsubsection*{Limited Function Scope}
In our prototype evaluation, we evaluate a CPU-intensive and a memory-intensive function.
The functions present in modern applications however often combine these attributes, and incorporate other tasks such as I/O and network traffic.

Our results however showed that our approach works with both types, so we do not suspect that the results would differ heavily. 
Preliminary experiments have been run with more complex functions, and similar results as presented in this paper were achieved. We aim to publish these in future work, along with further extending the implemented function types to provide a broader comparison ground and further validate our approach.

\subsubsection*{Choice of Deployment Approach}
We believe that our implemented choice is the most realistic environment that we can provide for duet benchmarking FaaS releases.

Early research included the use of a cloud instance, where we hosted each artifact in a separate Firecracker microVM, isolated them to only use one of the two vCPUs, and then called them in parallel.
Each of these containers hosted one release each, which were called by using AWS-provided instructions to build one's own Lambda functions outside the actual service.
This method had inherent benefits over a local deployment and our current approach, as we are able to more easily retain control over the executing instance.
We also do not have to rely on an external repository here, which could introduce other benefits.
The downside of this approach was that we still only emulated a FaaS environment, and did not actually use one.

Future work plans to compare both ideas and find benefits that either approach may have over the other.

\subsubsection*{Performance Change Detection}
We rely on Bootstrapping Percentile Intervals to determine a change in performance.
This form of analysis is currently state-of-the-art, however other approaches exist that we could have included and plan on including in future work~\cite{daly2020use,grambow2019continuous,toslali2021iter8}.
\section{Conclusion}
\label{sec:conclusion}

In this paper, we presented DuetFaaS, a duet benchmarking approach for continuous FaaS software application releases.
We evaluated our approach by comparing it to the traditional independent benchmarking strategy and randomized sequential (RMIT) methodology.

Our experiments with a proof-of-concept prototype on AWS Lambda showed that duet benchmarking improves benchmark accuracy when compared to the traditional and randomized sequential approach, and provided equal or more narrow confidence intervals with fewer results than other approaches in $98.41\%$ of evaluated cases.
When including only 100 results, duet benchmarking reduces the interval size by up to $99.28\%$, which helps developers better evaluate their function artifacts quickly.
We propose to include duet benchmarking for FaaS application releases, and plan to continue our work to improve the approach.


\balance

\bibliographystyle{ACM-Reference-Format}
\bibliography{bibliography.bib}

\end{document}